\documentclass[a4paper]{article}

\usepackage{INTERSPEECH2022}

\usepackage{amsmath,graphicx}

\usepackage[ruled,vlined]{algorithm2e}
\usepackage{booktabs}
\usepackage[flushleft]{threeparttable}
\SetKwComment{Comment}{$\triangleright$\ }{}

\newtheorem{definition}{Definition}[section]
\usepackage{bbm}

\title{Private Language Model Adaptation for Speech Recognition}
\name{Zhe Liu, Ke Li, Shreyan Bakshi, Fuchun Peng}
\address{Meta AI, Menlo Park, CA, USA}
\email{\{zheliu, kli26, shreyanb, fuchunpeng\}@fb.com}

\begin{document}

\maketitle
\begin{abstract}
Speech model adaptation is crucial to handle the discrepancy between server-side proxy training data and actual data received on local devices of users. With the use of federated learning (FL), we introduce an efficient approach on continuously adapting neural network language models (NNLMs) on private devices with applications on automatic speech recognition (ASR). To address the potential speech transcription errors in the on-device training corpus, we perform empirical studies on comparing various strategies of leveraging token-level confidence scores to improve the NNLM quality in the FL settings. Experiments show that compared with no model adaptation, the proposed method achieves relative 2.6\% and 10.8\% word error rate (WER) reductions on two speech evaluation datasets, respectively. We also provide analysis in evaluating privacy guarantees of our presented procedure.
\end{abstract}

\section{Introduction}
\label{sec:intro}

Neural network language models (NNLMs) play critical roles in automatic speech recognition (ASR) systems \cite{mikolov2010recurrent,chen2015improving,xu2018neural,irie2019language}. They typically outperform traditional $n$-gram LMs with better capability of modeling long-range dependency. For conventional ASR models, NNLMs are widely used in the second pass via $N$-best or lattice rescoring \cite{liu2014efficient,xu2018pruned,li2021parallelizable}. For end-to-end ASR \cite{graves2006connectionist,graves2012sequence,chan2016listen}, although linguistic information is implicitly learned, NNLMs can still further improve accuracy by fusion in first-pass decoding \cite{kannan2018analysis, kim2021improved} or second-pass rescoring.

With the latest advances in mobile technologies, hosting an ASR system entirely on-device has important implications from a reliability, latency, and particularly \emph{privacy} perspective, and has become an active area of research and industrial applications \cite{he2019streaming}. A common issue arising after deploying an ASR model on user devices is the discrepancy between training data and actual data received on local devices. The semantic and acoustic characteristics of real users' speech could be very different from those of server-side proxy data, in which case speech model \emph{adaptation} is indispensable. The privacy-preserving constraint requires user data to stay on their local devices. It is thus more challenging to perform model adaptation on user devices since there is no ground truth speech transcription from users.

To resolve this privacy concern, \emph{federated learning} (FL) \cite{konevcny2016federated, konevcny2016federated2, mcmahan2017communication}, a distributed learning technique, has been proposed and applied in many fields including recommendation \cite{chen2018federated}, keyboard suggestion \cite{arnold2016suggesting, ji2019learning}, keyword spotting \cite{leroy2019federated}, health care \cite{xu2019federated}, and more recently, ASR including hybrid acoustic models and end-to-end models \cite{dimitriadis2020federated, guliani2021training, cui2021federated}. FL can protect data privacy by training a shared model in a decentralized manner on users' local devices, so that raw data never leaves physical devices. Specifically, FL distributes the training process among a large number of client devices, with each client learning from live data and calculating model updates independently, then uploading those updates to a central server for aggregation. The updated model will later be delivered to each client device, after which this procedure is repeated until the training convergence of the  model.

In this work, we focus on federated NNLM adaptation for speech recognition application. Federated language modeling has been well explored in mobile keyboard suggestion where sentences typed by users provide instant labeled data for supervised learning \cite{ji2019learning}. However, for any on-device ASR  with privacy-preserving requirement, users' text data can not be directly accessed. Instead, we could use decoded hypotheses to perform model adaptation. The adaptation quality can be affected by any ASR transcription errors. Thus, more advanced methods are called for to better leverage transcribed data to conduct FL-based adaptation in an unsupervised manner.

To alleviate the transcription errors issue described above, we leverage confidence scores of transcripts, which estimate how likely each token in any decoded hypothesis from ASR is correct \cite{jiang2005confidence, huang2013predicting}. Lattice posteriors from conventional ASR systems can be directly used as confidence scores. Modeling based approaches, for example, confidence classifiers trained with various decoding features \cite{kalgaonkar2015estimating}, can provide more accurate confidence measurements. In this paper, we propose to mitigate errors in decoded hypotheses by modifying NNLM training objective using token-level confidence scores from a confidence classifier directly and improve adaptation quality.

The prior work on using ASR confidence scores in LM task is limited. Authors in \cite{xie2013evaluating} use confidence scores for selecting text data for LM adaptation. Our paper presents and investigates this direction in a rigorous way, proposes the weighting method for adjusting the cross-entropy loss, and conducts solid comparisons on the performance of these weighting approaches in the FL framework.

We mainly pursue three goals: (1) to present an effective procedure on FL-based domain adaptation of NNLMs with its applications on ASR; (2) to empirically compare approaches of using token-level confidence scores to improve adaptation quality; and (3) to provide analysis in evaluating privacy guarantees of our proposed method using \emph{differential privacy} (DP) tools \cite{dwork2006calibrating, dwork2014algorithmic}. To the best of our knowledge, our paper is the first work that leverages FL to fine-tune NNLMs for ASR systems and utilizes confidence scores to address any potential quality degradation due to mis-transcribed text as training corpus. In particular, the proposed confidence-based approach can also be applied to other tasks as well, for example, unsupervised speaker adaptation.

The rest of the paper is organized as follows. In Section~\ref{sec:method}, we introduce the FL-based domain adaptation approach of NNLMs with its applications on speech recognition tasks. We evaluate the proposed method in Section~\ref{sec:expt} and conclude in Section~\ref{sec:summary}.

\section{Methods}
\label{sec:method}
\subsection{Federated Adaptation Framework}
FL distributes the model training process across a large number of client devices. Each device trains on private data and computes model updates independently. Those updates are then uploaded to a central server for aggregation and the updated model is deployed to each client afterwards. We describe our approach on FL-based NNLMs adaptation for ASR as follows.

\emph{Pre-training}. We train an initial NNLM using a large general corpus and if available, any ``close-in-domain'' proxy data on the server side. The model then is delivered to each local device along with an ASR model;

\emph{Client-side update}. As soon as a user has input sufficient volume of utterances, local personal transcribed data is used to fine-tune all parameters of the current NNLM on the device. Then the client model update is sent back to the server if the device is selected to join the cohort;

\emph{Server-side update}. Once adequate client model updates are received by the server, global model update is conducted and an updated server model is deployed to each local device.

The client-side and server-side updates above are repeated until model convergence. The procedure is outlined in Algorithm~\ref{algo}, with more details in subsections \ref{method:client}, \ref{method:client:cclm}, and \ref{method:server}.

\begin{algorithm}
\SetAlgoLined
 Hyper-parameters $K, \eta_l, \eta_g, \beta_1, \beta_2, \epsilon_g$\;
 Initialize $\theta_1$ as a pre-trained NNLM (no adaptation)\;
 \For{\emph{each round} $t=1,2,\ldots$}{
  Deliver $\theta_t$ to each client\\
  Sample a subset $\mathcal{I}_t$ of clients\\
  \For{\emph{each client} $i\in\mathcal{I}_t$ \emph{in parallel}}{
    $\theta_{i,1}^t:=\theta_t$\\
    Load ASR transcripts on client $i$ for training\\
    \For{\emph{each local epoch} $k=1,2,\ldots, K$}{
        Compute gradients $g_{i,k}^t$ on batches\\
        $\theta_{i,k+1}^t\leftarrow\texttt{SGD}(\theta_{i,k}^t, g_{i,k}^t, \eta_l)$\\
    }
    $\theta_i^t:=\theta_{i,K+1}^t$\\
    Send $\Delta_i^t:=\theta_t-\theta_i^t$ to server\\
  }
  $\theta_{t+1}\leftarrow\texttt{FedAdam}(\theta_t, \Delta_i^t, w_i^t, t, \eta_g, \beta_1, \beta_2, \epsilon_g)$\\
 }
 \caption{FL-based NNLM adaptation for ASR.}
 \label{algo}
\end{algorithm}

\subsection{Client-side NNLM Adaptation}
\label{method:client}
Upon the initial deployment, on-device ASR model with the pre-trained NNLM for second-pass rescoring runs on each local client to transcribe user's utterances. The decoded hypotheses are then stored on the local device and serve as in-domain data for on-device NNLM adaptation. In this work, only the 1-best hypothesis of each utterance is stored and used for on-device training.

Suppose at round $t$ of FL training, each selected client downloads the NNLM $\theta_t$ from server and performs secure local training on their own device, that is, fine-tuning $\theta_t$ using private data. Mini-batch stochastic gradient descent (SGD) is used as the local optimizer. Specifically, in the $k$th training epoch, transcripts data on client $i$ is split into multiple batches. For the $b$th batch, let us denote $\theta_{i,k,b}^t$ as the current client model and $g_{i,k,b}^t$ as the computed gradients after back-propagation. Then the client model is updated as
\begin{align}
    \theta_{i,k,b+1}^t=\theta_{i,k,b}^t - \eta_l \cdot g_{i,k,b}^t(\theta_{i,k,b}^t)
\end{align}
where $\eta_l$ is a local learning rate. After $K$ epochs of training, the client uploads its model update (i.e.~difference of model parameters; see subsection \ref{method:server}) to the central server over a secure connection.

In the next subsection, we describe how to compute $g_{i,k,b}^t$ in speech LM task, and how to utilize token-level confidence scores to address the potential quality degradation due to mis-transcribed text as training data.

\subsection{NNLM Adaptation with Confidence Scores}
\label{method:client:cclm}
Cross entropy loss is usually used for NNLM training. The following shows this function for the $b$th batch of $k$th local training epoch on client $i$ 
\begin{align}
    \mathcal{L}_{i,k,b}^{t}(\theta)=-\frac{1}{n_b}\sum_{j=1}^{n_b}\frac{1}{T_j}\sum_{s=1}^{T_j}\log(\hat{p}_{j, s, v_{j,s}^*}(\theta))
\end{align}
where $n_b$ denotes the batch size, $T_j$ refers to the sequence length, $v_{j,s}^*$ represents the $s$th word of the $j$th transcript, and $\hat{p}_{j, s, v_{j,s}^*}$ indicates the predicted probability of observing $v_{j,s}^*$ over the entire vocabulary.

Adapting NNLMs using ASR transcribed data has the challenge of dealing with potential transcription errors. In this work, we leverage external confidence classifier models to mitigate this issue. Specifically, we modify the NNLM training objective using confidence scores. Let $\hat{c}_{j, s}$ be the estimated confidence score on the word of $v_{j,s}^*$ in the $j$th transcript, and 
\begin{align}
    {\hat{c}}_{j}:=\frac{1}{T_j}\sum_{s=1}^{T_j}\hat{c}_{j, s}
\end{align}
be the estimated utterance-level confidence score, which is the average of token-level confidence scores. We propose the following three modified loss functions for NNLM training.

\textbf{Hard thresholding}. We adopt utterance-level confidence scores for training data selection, which amounts to exclude the set of utterances $$\{j\in[n_b]: \hat{c}_{j}<c\}$$ from training, where $c$ is some fixed constant. Notice that this method is equivalent to include an indicator function as a multiplier on each utterance in the loss function.

\textbf{Utterance-level weighting}. The utterance-level confidence scores are leveraged for loss weighting
\begin{align}
    \mathcal{L}_{i,k,b}^{t,\text{utt-weight}}(\theta)=-\frac{1}{n_b}\sum_{j=1}^{n_b}\frac{\hat{c}_{j}}{T_j}\sum_{s=1}^{T_j}\log(\hat{p}_{j, s, v_{j,s}^*}(\theta))
\end{align}

\textbf{Token-level weighting}. We utilize token-level confidence scores for weighting in the loss function
\begin{align}
    \mathcal{L}_{i,k,b}^{t,\text{token-weight}}(\theta)=-\frac{1}{n_b}\sum_{j=1}^{n_b}\frac{1}{T_j}\sum_{s=1}^{T_j}\hat{c}_{j, s}\log(\hat{p}_{j, s, v_{j,s}^*}(\theta))
\end{align}

\subsection{Server-side NNLM Update}
\label{method:server}
Suppose that at round $t$, the server has the model $\theta_t$ and samples a set $\mathcal{I}_t$ of clients. Let $\theta_i^t$ denotes the model of each client $i\in \mathcal{I}_t$ after local training, and $\Delta_i^t:=\theta_t-\theta_i^t$ be the model difference on client $i$ which is sent back to the server. Let
\begin{align}
    \Delta_t:=\frac{\sum_{i\in\mathcal{I}}w_i^t\Delta_i^t}{\sum_{i\in\mathcal{I}} w_i^t}
\end{align}
be the averaged model difference or ``pseudo-gradient'' which is used in general server optimizer
updates. Here, $w_i^t$ refers to the weight being assigned to the model difference from client $i$ in the aggregation, i.e. number of words in the training data for adapting the client NNLM in round $t$.

We use the \texttt{FedAdam} optimizer for updating the global model \cite{reddi2020adaptive}. Specifically, let $\eta_g$ be the learning rate and hyper-parameters $\beta_1$, $\beta_2\in[0, 1)$, then
\begin{gather}
    m_{t} = \beta_1 m_{t-1}+(1-\beta_1) \Delta_t \\ 
    v_{t} = \beta_2 v_{t-1}+(1-\beta_2) (\Delta_t)^2 \\ 
    \hat{m}_{t} = m_{t}/(1-\beta_1^t) \\
    \hat{v}_{t} = v_{t}/(1-\beta_2^t) \\
    \theta_{t+1} = \theta_{t}-\eta_g\cdot\frac{\hat{m}_t}{\sqrt{\hat{v}_t}+\epsilon_g}
\end{gather}
where $m_0=0, v_0=0$, and $\epsilon_g$ is a small positive number.

\subsection{Differential Privacy for NNLM Adaptation}
\label{method:dp}

A differentially private mechanism enables the public release of model parameters with a strong privacy protection \cite{dwork2006calibrating, dwork2014algorithmic}.
\begin{definition}[DP]
\label{def:dp}
A randomized mechanism $\mathcal{M}$ with a domain $\mathcal{D}$ and range $\mathcal{S}$ satisfies $(\epsilon,\delta)$-DP if for any two adjacent datasets $d,d'\in\mathcal{D}$ and for any subset $S\subseteq\mathcal{S}$, it holds that
\begin{align}
    P(\mathcal{M}(d)\in S)\leq e^\epsilon P(\mathcal{M}(d')\in S) + \delta.
\end{align}
\end{definition}
Here $d$ and $d'$ are defined to be \emph{adjacent} if $d'$ can be formed by adding or removing a single training example from $d$. It is worth noting that the definition of adjacent datasets in Definition~\ref{def:dp} depends on the application. Most prior work on DP deals with example level (or utterance level in our case). For our task, a better definition is user-level adjacency for protecting whole user histories in the training set \cite{brendan2018learning}, since a sensitive word may be uttered several times by an individual user. Note that given some target $\delta$, a smaller $\epsilon$ leads to stronger privacy protection, but often, can degrade the model accuracy.

Two modifications to the FL-based NNLM adaptation are needed to ensure an differentially private algorithm. First, clip the gradient computed on any client per each round to bound a user's impact on model parameters. Second, add randomly sampled \emph{Gaussian} noise to the clipped gradient.

In subsection \ref{expt:privacy}, we perform privacy analysis of the proposed NNLM adaptation approach.

\section{Experiments}
\label{sec:expt}

\subsection{Datasets}
\label{expt:data}
In our experiments, the ASR model is trained using in-house video dataset (14K hours), which is sampled from public social media videos and de-identified before transcription; both transcribers and researchers do not have access to any user-identifiable information (UII).

The following two ASR applications are considered. The first is \emph{conversation} speech and the second is short \emph{voice command} for smart devices. For both use cases, in-domain speech datasets are collected using mobile devices through crowd-sourcing from a data supplier for ASR; the data is properly anonymized and no UII is contained in the datasets.

Table~\ref{tab:data} shows a summary of the two datasets which contain the \texttt{adaptation} split for NNLMs adaptation, and the \texttt{test} split for model evaluation purpose. It is worth mentioning that the word error rates (WERs) on the \texttt{adaptation} split using the video ASR model with a pre-trained NNLM for second-pass rescoring are 9.7\% and 12.2\% for \emph{conversation} and \emph{voice command} applications, respectively.

\begin{table}[ht]
  \caption{Summary of speech datasets for two applications.}
  \centering
  \resizebox{\columnwidth}{!}{%
  \begin{tabular}{ll|r|r}
    \toprule
    \emph{Split} & \emph{Feature}& \emph{Conversation} & \emph{Voice Command}  \\
    \midrule
    \texttt{adaptation} & \# of utts & 166K & 63K \\
    & \# of words & 1,738K & 363K \\
    \midrule
    \texttt{test} & \# of utts & 13K & 13K \\
    & \# of words & 123K & 73K \\
    \bottomrule
  \end{tabular}
  }
  \label{tab:data}
\end{table}

\subsection{Setups}
\label{expt:setups}
We would like to simulate the real-world scenarios after deploying the ASR and pre-trained NNLM models to clients. Voice data from the \texttt{adaptation} split is streamed to each device and decoded by the on-device speech models. Then the transcripts are used to fine-tune the NNLMs for domain adaptation.

For the ASR model, we use connectionist temporal classification (CTC) \cite{graves2006connectionist} criterion to learn an acoustic model and it is further composed with a 5-gram LM in a standard weighted finite-state transducers framework. Here, we adopt a 6-layer latency-control bi-directional LSTM encoder with hidden dimension 1,000. The NNLM is used for second-pass 5-best rescoring, where we utilize a LSTM based model with character embeddings \cite{kim2016character} dimension 100, and 2 layers of 512 hidden units. For each hypothesis among the 5-best list, its NNLM score is linearly interpolated with the score from the 5-gram LM. The interpolation weight is chosen to give the lowest WER. The confidence classifier model is trained on video dataset using feed-forward networks and handcrafted input features from decoding results; isotonic regression is used for calibrating the model.

\begin{figure}[t]
  \centering
  \includegraphics[width=\linewidth]{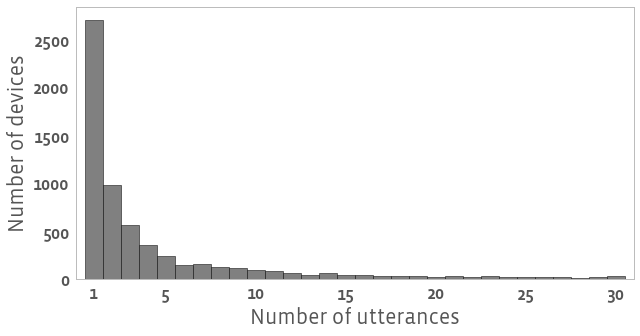}
  \caption{Histogram of number of utterances on each device in the simulation.}
  \label{fig:hist}
\end{figure}

To simulate the environment of FL-based approaches, for each utterance, we generate a random device label from a \emph{Zipf} distribution. Thus utterances with a common device label are considered being received by the same device. This results in approximately 8K devices for each application. Figure~\ref{fig:hist} shows the histogram of number of training examples (i.e. utterances) on each device in the simulation, where we can see the empirical distribution is highly skewed.

Regarding the hyper-parameters of FL training, we set the number of selected users per round $|\mathcal{I}_t|=100$; learning rate $\eta_g=0.001$ in the global \texttt{FedAdam} optimizer and $\eta_l=1.0$ for the client SGD optimizer.  Locally, we only train 1 epoch with batch size 8 for any selected client per each FL round. We use 10 epochs for FL training, where each epoch corresponds to 100 rounds.

\subsection{Evaluation Results}
\label{expt:res}

The baseline model in our experiments is the one where we use the on-server pre-trained NNLM for rescoring, without domain adaptation. We compare it with the fine-tuned NNLM using in-domain unsupervised text from the \texttt{adaptation} split (transcribed by the ASR model) in the FL frameworks. Note that such in-domain data never leaves physical devices and is thus not accessible from servers due to privacy restrictions. Multiple methods in handling potential transcription errors in the adaptation data are measured, including using all transcripts, hard thresholding, utterance-level and token-level weighting. For comparison purposes, we also include the result without NNLM rescoring.

Table~\ref{tab:eval:conv} and Table~\ref{tab:eval:voice} show the perplexity (PPL) and WER results on the \emph{conversation} and \emph{voice command} evaluation datasets. Compared with the baseline model, FL-based domain adapted NNLMs (using all transcripts) obtain relative 2.1\% and 8.4\% WER gains on the two use cases. In addition, models leveraging confidence scoring always outperform the one using all transcripts as the training data, and obtain up to relative 0.5\% and 2.4\% WER reductions on the two applications, respectively. For short \emph{voice command} evaluation set, hard thresholding leads to the best result. For longer \emph{conversation} utterances, token-level weighting performs the best.

\begin{table}[ht]
  \caption{Results on the Conversation evaluation set.}
  \centering
  \resizebox{\columnwidth}{!}{%
  \begin{threeparttable}
  \begin{tabular}{l|r|l}
    \toprule
    & \multicolumn{2}{|c}{\emph{Conversation}} \\
    \cmidrule(r){2-3}
    \emph{Model}& \emph{PPL} & \;\emph{WER}  \\
    \midrule
    No NNLM & -\,\,\,\, & \;\;8.07\;\; \\
    \midrule
    Server-pretrained NNLM (no adapt) & 109.4 & \;\;7.96\;\; \\
    \midrule
    FL-finetuned NNLM (all utts) & 32.0 & \;\;7.79 (-2.1\%)\;\; \\
    FL-finetuned NNLM (hard thres. utts) & 31.2 & \;\;7.76 (-2.5\%)\;\; \\
    FL-finetuned NNLM (utt weighted) & 30.8 & \;\;7.77 (-2.4\%)\;\; \\
    FL-finetuned NNLM (token weighted) & 30.3 & \;\;\bf{7.75} (-2.6\%)\;\; \\
    \bottomrule
  \end{tabular}
  \end{threeparttable}
  }
  \label{tab:eval:conv}
\end{table}

\begin{table}[ht]
  \caption{Results on the Voice Command evaluation set.}
  \centering
  \resizebox{\columnwidth}{!}{%
  \begin{threeparttable}
  \begin{tabular}{l|r|l}
    \toprule
    & \multicolumn{2}{|c}{\emph{Voice Command}} \\
    \cmidrule(r){2-3}
    \emph{Model}& \emph{PPL} & \;\emph{WER}  \\
    \midrule
    No NNLM & -\,\,\, & 10.10 \\
    \midrule
    Server-pretrained NNLM (no adapt) & 420.4 & \;\;9.89 \\
    \midrule
    FL-finetuned NNLM (all utts) & 8.1 & \;\;9.06 (-8.4\%) \\
    FL-finetuned NNLM (hard thres. utts) & 8.0 & \;\;\bf{8.82} (-10.8\%) \\
    FL-finetuned NNLM (utt weighted) & 8.0 & \;\;9.00 (-9.0\%) \\
    FL-finetuned NNLM (token weighted) & 8.0 & \;\;8.93 (-9.7\%) \\
    \bottomrule
  \end{tabular}
  \end{threeparttable}
  }
  \label{tab:eval:voice}
\end{table}

We also evaluate the performance of NNLM adaptation without pre-training, that is, training from scratch using the in-domain data in the FL settings. From the results in Table~\ref{tab:eval:conv:scratch}, we can see that training from scratch has around relative 1\% WER degradation comparing to fine-tuned models in the FL settings. Thus it is beneficial to start with some pre-trained NNLM before on-device adaptation.

\begin{table}[ht]
  \caption{Results on the Conversation evaluation set without pre-training on NNLMs.}
  \centering
  \resizebox{\columnwidth}{!}{%
  \begin{threeparttable}
  \begin{tabular}{l|c|l}
    \toprule
    & \multicolumn{2}{|c}{\emph{Conversation}} \\
    \cmidrule(r){2-3}
    \emph{Model}& \emph{PPL} & \emph{WER} \\
    \midrule
    FL-finetuned NNLM (all utts) & 32.0 & 7.79 \\
    FL-finetuned NNLM (token weighted) & 30.3 & 7.75 \\
    \midrule
    FL-trained-from-scratch NNLM (all utts) & 35.8 & 7.85 \\
    FL-trained-from-scratch NNLM (token weighted) & 34.5 & 7.82 \\
    \bottomrule
  \end{tabular}
  \end{threeparttable}
  }
  \label{tab:eval:conv:scratch}
\end{table}

\subsection{Privacy Analysis}
\label{expt:privacy}

Our privacy analysis is performed in the framework of Rényi DP \cite{mironov2017renyi}, which is a natural relaxation of DP based on the Rényi divergence. It is well-suited for expressing guarantees of privacy-preserving approaches and for composition of heterogeneous mechanisms.

In this analysis, the $L_2$ norm clip is set to 0.5, and the noise multiplier, which is the ratio of the standard deviation of Gaussian noise to the $L_2$ sensitivity, is set to 0.2, 0.5, and 1.5 in our experiments. We set the target $\delta$ as 1e-5 and the value of $\epsilon$ is calculated accordingly.

Table~\ref{tab:eval:conv:privacy} displays the privacy analysis results on FL-based adapted NNLMs using all or token-level weighted transcripts. It is expected that the lower the value of $\epsilon$, the larger the PPL and WER. It is worth noting that these resulting models still perform better than the server-trained NNLM without domain adaptation, although the margins of gains become smaller than the one without strong privacy guarantees.

\begin{table}[ht]
  \caption{Privacy analysis on the Conversation evaluation set.}
  \centering
  \resizebox{\columnwidth}{!}{%
  \begin{threeparttable}
  \begin{tabular}{l|r|c|l}
    \toprule
    & \multicolumn{3}{|c}{\emph{Conversation}} \\
    \cmidrule(r){2-4}
    \emph{Model} & $(\epsilon, \delta)$-\emph{DP}\,\,\, & \emph{PPL} & \emph{WER} \\
    \midrule
    & (248.6, 1e-5) & 46.1 & 7.88 \\
    FL-finetuned NNLM (all utts) & (14.2, 1e-5) & 50.4 & 7.91 \\
    & (0.9, 1e-5) & 56.7 & 7.92 \\
    \midrule
    & (248.6, 1e-5) & 45.9 & 7.85 \\
    FL-finetuned NNLM (token weighted) & (14.2, 1e-5) & 50.4 & 7.89 \\
    & (0.9, 1e-5) & 56.3 & 7.93 \\
    \bottomrule
  \end{tabular}
  \end{threeparttable}
  }
  \label{tab:eval:conv:privacy}
\end{table}

\section{Conclusion}
\label{sec:summary}
In this paper, we introduce a NNLM adaptation approach for ASR in the FL settings. Particularly, we leverage confidence scoring models to adjust the NNLM training objective accordingly. Experiments show that compared with no adaptation, the presented method obtains modest WER reductions on two speech datasets. We also perform privacy analysis of the proposed approach using DP. Future work includes exploring the personalization of NNLMs in a FL framework. 

\bibliographystyle{IEEEtran}
\bibliography{refs}

\end{document}